\documentstyle[aps,twocolumn,epsfig,amsfonts]{revtex}

\begin{document}
\title{Mean field effects in a trapped classical gas}

\author{D. Gu\'ery-Odelin}

\address{Laboratoire Kastler Brossel$^*$, Ecole normale sup\'erieure}
\address{24, Rue Lhomond, F-75231 Paris Cedex 05, France}

\date{\today}

\maketitle

\begin{abstract}
In this article, we investigate mean field effects for a bosonic
gas harmonically trapped above the transition temperature in the
collisionless regime. We point out that those effects can play
also a role in low dimensional system. Our treatment relies on the
Boltzmann equation with the inclusion of the mean field term.
 The equilibrium state is first discussed. The dispersion relation for collective
oscillations (monopole, quadrupole, dipole modes) is then derived.
In particular, our treatment gives the frequency of the monopole
mode in an isotropic and harmonic trap in the presence of mean
field in all dimensions.
\end{abstract}


The dynamics of Bose-Einstein condensates (BEC) of dilute atomic
gases are described by the Gross-Pitaevskii equation (\cite{revue}
and references therein). The main feature of this equation is the
mean-field term arising from interaction between particles. Most
of BEC experiments are in the so-called Thomas-Fermi regime for
which the interaction energy dominates the kinetic energy,
resulting in an inverted parabola density shape of the condensate.
However, the mean-field term is not found only in a bosonic gas
well below its critical temperature $T_c$. Such a contribution
from collisions also exists above $T_c$, and is even magnified by
a kind of Hanbury-Brown and Twiss factor. Up to now most BEC
experiments have  been performed in the collisionless regime (when
the mean free collision rate is small relative to the trap
frequency) and with a negligible contribution of the mean-field of
the non-condensed atoms.

 In order to specify the role of dimensionality, we introduce the
 adimensional parameter $\zeta=gn/k_BT$ {\it i.e.} the ratio
 between the mean field energy and the thermal energy. One
 readily establishes that $\zeta_{3D}\sim(n
 a^3)^{1/3}(n\lambda^3_{dB})^{2/3}$, where $\lambda_{dB}=h(2\pi mk_BT)^{-1/2}$ is the de Broglie
  wavelength and $a$ the $s$-wave scattering length. Consequently for a dilute bosonic gas above
 the critical temperature: $\zeta_{3D}\ll 1$ and results
  presented in this paper are valid as corrections. In the weakly interacting limit, the 2D quantity $\zeta_{2D}$
 is only  logarithmically small
 with respect to $\lambda^{2}_{dB}$
 and mean field energy can be comparable to thermal energy above
 the quantum transition temperature (Kostherlitz-Thouless \cite{b2D}). The quantity $\zeta_{1D}$
 is of the order of $
 (n\lambda_{dB})^2(nl_c)^{-2}$ where the correlation length $l_c=\hbar/\sqrt{mgn}$.
Classical description and mean field can be used up to the regime
where $\lambda_{dB}\sim 1/n\sim l_c$ so up to $\zeta_{1D}\sim 1$.
In the
  regime where $l_c$
 is much smaller than the mean interparticle separation
 the gas acquires Fermi properties and is called a gas of impenetrable bosons
 or Tonk gas \cite{b1D}.

 Finally, by reduction of the dimensionality the role
 of the mean field even above the critical temperature of a quantum transition
 can be important. Experiments performed
  on microchip offer the possibility to investigate low
  dimensional regime \cite{Jacob,Zimmermann} as well as experience
  with dipolar trap or/and magnetic trap\cite{dipolar}.

In this paper, our aim is to extend the traditional treatment of
the bosonic gas above the critical temperature by taking into
account the effect of particle interactions. The method consists
in including
 the classical mean field term, also known as the Vlasov
contribution, within the Boltzmann equation. So far, collective
oscillations of a bosonic gas above the critical temperature have
been investigated without mean field contribution in the
hydrodynamic regime in Ref. \cite{Griffin,shlyap}, and an
interpolation formula from the collisionless up to the
hydrodynamic regime has been proposed in Ref.
\cite{dgo99,khawaja}.

In Sec. I, we briefly recall the general framework that describes
how the mean field is taken into account in the Boltzmann
equation. The stationary solution is discussed in Sec. II. In Sec.
III, we derive equations of the low energy collective excitation
of a Bose gas for positive and negative scattering length by means
of a scaling ansatz. We report an interpolation formula from the
collisionless gas to the interaction dominated thermal gas (Vlasov
gas) in the absence of dissipation.

\section{formulation}

In traditional BEC experiments, the bosonic gas above the critical
temperature is well-described by the classical Boltzmann equation
\cite{Huang}, or the Uhlenbeck-Boltzmann equation \cite{BaymLivre}
if experiments are sufficiently accurate to measure deviation from
the classical distribution. In the following, we present  the
extension of this equation when the mean-field is taken into
account.

We consider an ensemble of harmonically trapped thermal atoms
 that evolves according to the
Boltzmann-Vlasov kinetic equation\cite{Huang,Kreuzer}:
\begin{equation}
\frac{\partial f}{\partial t}+{\bf v}\cdot\frac{\partial
f}{\partial{\bf r}}-\frac{\partial U}{\partial {\bf
r}}\cdot\frac{\partial f}{\partial {\bf v}}
-2\frac{g}{m}\frac{\partial n}{\partial {\bf
r}}\cdot\frac{\partial f}{\partial {\bf v}}=I_{\rm coll}
\label{BV}
\end{equation}
where $f({\bf r},{\bf v},t)$ is the single particle phase space
distribution function, $U=\sum_i\omega_i^2r_i^2/2$ the confining
potential, $n=\int fd^Dv$ the density, where $g$ is the mean field
strength in $D$ dimensions. In 3D, $g=4\pi\hbar^2a/m$ is the
strength of the pseudopotential replacing the true two-body
potential at low energies , with the $s$-wave scattering length
$a$. $I_{\rm coll}$ is the collisional integral that describes
relaxation processes. Note that $I_{\rm coll}=0$ in 1D because of
conserved quantities. The Vlasov term (last term of the l.h.s. of
(\ref{BV})) is a Hartree-Fock mean-field term \cite{YvanHouches}
and is even magnified for point-like interactions by a factor of
two with respect to the condensate for the same density. Indeed,
for non-condensed cloud both the Hartree and the Fock terms
contribute, whereas only the Hartree term contributes for the
condensate.

 The kinetic equation (\ref{BV}) is valid for $k_BT\gg
 \hbar\omega$ where $\omega$ is the typical trap frequency and for
 $a\ll a_0$ where $a_0=(\hbar/(m\omega))^{1/2}$ is the oscillator
  quantum length. One has to check that the $s$-wave approximation
  is valid. In 3D it requires that $a\ll \lambda_{dB}$.

\section{Equilibrium state}
At equilibrium, the Eq. (\ref{BV}) reads:
\begin{equation}
\sum_{i=1}^D\bigg(v_i\frac{\partial f_0}{\partial
r_i}-\omega_i^2r_i\frac{\partial f_0}{\partial
{v_i}}-\frac{2g}{m}\frac{\partial n_0}{\partial
r_i}.\frac{\partial f_0}{\partial {v_i}}\bigg)=0. \label{eq}
\end{equation}
By multiplying Eq. (\ref{eq}) by $v_jr_j$ and integrating over
space and velocity, we deduce the average size $\langle
r_j^2\rangle$ along the $j$ axis \cite{dgo99}:
\begin{equation}
\omega_j^2\langle r_j^2\rangle-\langle
v_j^2\rangle-\frac{g}{mN}\int n_0^2d^Dr=0.
\end{equation}
As expected, repulsive interactions ($g>0$) favor a reduction of
the density from the free particle situation. The opposite
behavior is obtained in case of attractive interactions ($g<0$).
One can extract the shape of the density by searching for a
factorized solution of (\ref{eq}) of the form: $f_0({\bf r},{\bf
v})=n_v({\bf v})n_0({\bf r})$. One finds a Gaussian spherical
distribution for the velocity. The density distribution is a
solution of the following equation:
\begin{equation}
\kappa\mbox{ln}n_0+2gn_0/m=\mu-\sum_j\omega_j^2r_j^2/2,
\label{dens}
\end{equation}
where $\kappa=\int v_j^2n_v d^Dv=k_BT/m$. In two limiting cases,
the solution has a simple form. For $g=0$, we find the gaussian
shape as expected for an harmonic confinement without the Vlasov
term. On the contrary, in the limit in which the interparticle
interactions dominate and are repulsive, the shape of the cloud is
determined by a balance between the harmonic oscillator and
interactions energy resulting in approximately an inverted
parabola shape for repulsive interactions. This is the same shape
as found for a harmonically trapped BEC in the Thomas-Fermi regime
\cite{revue}, since in this case the mean-field term also
dominates. Note that this last result can also be shown in the
classical hydrodynamic regime \cite{shlyap} under the same
conditions. Strictly speaking this situation can be reached only
for low dimensional system. For intermediate $g$, Eq. (\ref{dens})
gives the proper interpolation between the gaussian and the
Thomas-Fermi shape.

 For $g<0$, the density distribution is sharpened with respect to the free gaussian one.
Actually, by increasing the number of atoms, the spatial extent of
the distribution is reduced. If attractive forces overwhelm the
kinetic energy, the cloud should collapse. One can work out the
criterium for such a collapse in 3D by means of a gaussian ansatz
\cite{baym96} and finds $a_c=33a_0 N^{-1}(a_0/\lambda_{dB})^5$.
However this result is out of the range of validity of the
classical approximation.

\section{Collective oscillations of a collisionless gas}

In this section, we investigate the collective oscillations of a
Vlasov gas {\it i.e.} in the absence of the dissipative term
($I_{\rm coll}$) but with the mean field contribution.

\subsection{Scaling ansatz method}

 We study those modes by means
 of the scaling factor method \cite{shlyap,dum,zoller,singh,stoof}
in $D$ dimensions. We recall that in this method
 the proper shape of the cloud does not enter directly in the
 equations. This is the reason why the solutions are equally
 valid for a Bose gas just above the critical temperature as
 for a classical gas.
We make the
 following ansatz for the non equilibrium distribution function:
 $f({\bf r},{\bf v},t)=f_0({\bf R}(t),{\bf V}(t))$ with
 $R_i=r_i/\lambda_i$ and $V_i=\lambda_iv_i-\dot{\lambda}_ir_i$.
 The dependence in $t$ is contained in the free parameters $\lambda_i$.
  By substituing this ansatz into Eq. (\ref{BV}), we find:
 \begin{eqnarray}
\sum_i\bigg\{ \frac{V_i}{\lambda_i^2}\frac{\partial f_0}{\partial
R_i}-\lambda_iR_i(\ddot{\lambda}_i+\omega_i^2\lambda_i)
\frac{\partial f_0}{\partial V_i}
\nonumber\\
-\frac{2g}{\Pi_j\lambda_j} \frac{\partial n_0}{\partial
R_i}\frac{\partial f_0}{\partial V_i} \bigg\}\simeq 0. \label{e16}
 \end{eqnarray}
 This equation can be combined with Eq. (\ref{eq}) taken
 in the phase space point $({\bf r}={\bf R},{\bf v}={\bf V})$ in
 order to replace the last term of (\ref{e16}) by a linear
 superposition of $\partial f_0/\partial R_i$ and $\partial f_0/\partial
 V_i$. We finally obtain:
\begin{eqnarray}
\sum_i\bigg\{
\bigg(\frac{V_i}{\lambda_i^2}-\frac{V_i}{\Pi_j\lambda_j}\bigg)\frac{\partial
f_0}{\partial R_i}\nonumber\\
-\lambda_iR_i\bigg(\ddot{\lambda}_i+\omega_i^2\lambda_i-\frac{\omega_i^2}{\lambda_i\Pi_j\lambda_j}\bigg)
\frac{\partial f_0}{\partial V_i} \bigg\}=0. \label{e19}
 \end{eqnarray}
This equation provides the constraints on our ansatz. The first
average moment\cite{dgo99} of $R_iV_i$, namely $\int R_iV_i[Eq.
(\ref{e19})] d^DRd^DV/N$, leads to a set of Newton-like second
order ordinary differential equations:
\begin{equation}
\ddot{\lambda}_i+\omega_i^2\lambda_i-\frac{\omega_i^2}{\lambda_i^3}
+\omega_i^2\xi\bigg( \frac{1}{\lambda_i^3}-
\frac{1}{\lambda_i\Pi_j\lambda_j}\bigg)=0 \label{res}
\end{equation}
with $\xi=g\langle n_0\rangle/(g\langle n_0\rangle+k_BT)$. To find
the excitation frequencies of the modes, one must linearize around
the equilibrium value: $\lambda_i=1$.

\subsection{Monopole mode}

Consider the case of an isotropic harmonic confinement
($\omega_i=\omega_0$). In this case the scaling ansatz is in the
kernel of the collision integral and provides a solution of
(\ref{BV}) valid also in the collisional regime. An exact solution
of this mode was first reported in \cite{Boltzmann} for the
classical Boltzmann equation without mean field.

The small amplitude expansion of (\ref{res}) gives the frequency
of the monopole mode (also called the breathing mode): $\omega_0
\sqrt{4+\xi(D-2)}$, for all dimensions and in presence of the
total effects of collisions (mean field term and dissipation {\it
via} $I_{\rm coll}$). In 3D this frequency ranges from $2\omega_0$
in the absence of the mean field term up to $\sqrt{5}\omega_0$
when the mean field dominates, in the latter case one obtains the
same result as expected for Bose-Einstein condensate in the
Thomas-Fermi limit \cite{sandro}. In 2D we find $2\omega_0$ for
the monopole mode a result independent of the mean field, a
special feature of 2D already investigated in Ref. \cite{rosch}.
In this case, the scaling ansatz provides an exact solution of
(\ref{BV}). In 1D, the monopole frequency ranges from $2\omega_0$
down to $\sqrt{3}\omega_0$ when the mean field dominates. Note
that this latter result gives exactly the same frequency as the
one of the monopole mode in trapped ions \cite{james}. For ions
the force originates from the coulomb interaction, this long range
force is well described by mean field and this is probably the
reason why we recover the same result.

\subsection{Quadrupolar mode}

In 2D and 3D, Eq. (\ref{res}) provides the mean field contribution
to quadrupolar  collective oscillations for a collisionless gas.

In 2D, Eq. (\ref{res}) gives two coupled equations for $\lambda_1$
and $\lambda_2$, which, after linearization, yield the dispersion
relation
\begin{eqnarray}
\omega^2=\frac{1}{2}\bigg[ (4-\xi)(\omega_x^2+\omega_y^2)\nonumber \\
  \pm
  ((4-\xi)^2(\omega_x^2+\omega_y^2)-32\omega_x^2\omega_y^2)^{1/2}\bigg].
  \label{res2}
\end{eqnarray}
For $\xi=0$, we recover the single particle excitation frequency
of the cloud : $\omega=2\omega_i$ for each spatial direction. In
the limit $\xi=1$, this relation can be derived from a purely
hydrodynamic approach \cite{Griffin,shlyap} by taking into account
the mean field contribution in the same limit. Formula
(\ref{res2}) provides also finite temperature corrections to this
regime and the proper interpolation in between those two limiting
cases.

For a cylindrical 3D harmonic trap (we denote
$\beta=\omega_z/\omega_\perp$) we find the eigenfrequencies of
mode $M=0$ (coupling between quadrupole and monopole modes):
\begin{eqnarray}
\omega^2=\frac{1}{2}\bigg[ 4+4\beta^2-\xi\nonumber \\
  \pm
  ([4+4\beta^2-\xi]^2+8\beta^2[-8+2\xi+\xi^2])^{1/2}\bigg],
\end{eqnarray}
 and the
frequency of the quadrupole mode with azimuthal quantum number
M=2: $\omega^2/\omega^2_\perp=2(2-\xi)$.
 The limit
$\xi\sim 1$ gives the formulas derived by Stringari \cite{sandro}
for the low energy excitation spectrum of a BEC in the limit for
which the energy of interaction predominates over the kinetic
energy. Low energetical collective excitations (monopole,
quadrupole) spectrum of a BEC are rather a proof of mean field
dominated physics than a direct proof of superfluidity in the
Landau sense \cite{Landau}. As already pointed out, the validity
of our calculation in 3D is only perturbative ($\xi \ll 1$).
 For an isotropic 3D
trap ($\omega_i=\omega_0$), the oscillation frequency split into
the monopole mode with a frequency $\omega_M\simeq 2\omega_0
(1+\xi/8)$ and the quadrupole
 mode with a frequency $\omega_Q\simeq 2\omega_0 (1-\xi)$ as soon as
 we take into account the mean field.

\subsection{Dipolar mode}

The dipolar mode which corresponds to the rigid motion of the
density profile is not affected by the mean field. This can be
shown by searching for a solution of the form $f({\bf r},{\bf
v},t)=f_0({\bf R}(t),{\bf V}(t))$ with $R_i=r_i-\eta_i$ and
$V_i=v_i-\dot{\eta}_i$. Each component $\eta_i$ is time dependent.
Following the same procedure, we readily establish the equation of
motion for $\eta$ by taking the average value of $V_i$:
$\ddot{\eta}_i+\omega_i^2\eta_i=0$. We recover the fact that Kohn
modes do not depend on interactions. This last result is naturally
unchanged if we take into account the collisional integral
contribution.

\subsection{Discussion}

Note that $I_{\rm coll}=0$ is strictly speaking only applicable
for a collisionless gas ($a\rightarrow 0$) and to the hydrodynamic
regime (collision rate $\gg$ trap frequencies).
 In between, $I_{\rm coll}$ is negligible with respect to the
Vlasov term if $a\ll \tilde{a}$, where $\tilde{a}$ is a critical
value for the scattering length. The collision integral is of the
same order of magnitude as the Vlasov contribution when $g=\sigma
v^2 \ell$, where $\sigma=8\pi a^2$ is the elastic cross section,
$v$ is a typical thermal velocity and $\ell$ a typical size. We
take $\ell\sim v/\omega$ and find $\tilde{a}=0.03 \lambda_{dB}
(\lambda_{dB}/a_0)^2$. Just above the critical temperature
$\tilde{a}$ is of the order of the scattering length for the
experiment of Ref. \cite{Jacob} performed on a microchip. For the
metastable helium experiment \cite{heliumbecens} $\tilde{a}$ is
slightly smaller than the scattering length. In these experiments
even if they are
 not in the Thomas-Fermi regime, one can no longer ignore mean field
effects. The interpolation parameter in Eq. (\ref{res}) is the
ratio between the mean field and the thermal energy:
$\zeta=gn/k_BT$. In many BEC experiments, $\zeta_{max} < 10^{-4}$
which clearly justifies that it is neglected. However, in Ref.
\cite{Jacob} this ratio is of the order of $\zeta_{max}\sim 10$\%
and could be increased by a stronger longitudinal confinement. In
\cite{heliumbecens}, this ratio is of the order of
$\zeta_{max}\sim 20\pm$10\% as a consequence of the huge value of
the scattering length and the high density of the sample.

\section{Conclusion}

Mean field effects for a bosonic gas above quantum transition play
an increasing role as the dimension is reduced. This paper deals
with the contribution of the mean field to the low energetical
collective modes of such a gas. We derive, even for a collisional
gas, the frequency of the monopole mode for an isotropic and
harmonic confinement.
 Note that results
derived in this
  article hold also for two-components
 Fermi-system as soon as the mean field play a role.

 The mean field contribution could be seen directly on the equilibrium shape of the gas {\it
in situ}. Time-of-flight measurement may give relevence to a
direct observation of mean field contribution \cite{arimondo}.

Further experiments on microchip \cite{Zimmermann} should allow
the reduction of dimensionality. Dipolar trap can also be a good
tool \cite{dipolar}. Those techniques should help for the
observation of classical mean field above a quantum transition.

\section*{acknowledgments}
I thank the ENS Laser cooling group, F. Laloe and G. Shlyapnikov
for very stimulating discussions.

$^*$ Unit\'e de Recherche de l'Ecole normale sup\'erieure et de
l'Universit\'e Pierre et Marie Curie, associ\'ee au CNRS.

\end{document}